\documentclass[jap,showpacs,preprint]{revtex4}

\usepackage{graphicx,amsmath,amsfonts,textcomp,array}

\newcommand{\fref}[1]{Fig.~\ref{#1}}

\newcommand{\de}{$^{\circ}$ }

\begin{document}
\title{Lateral piezoelectric response across ferroelectric domain walls in thin films}
\author{J. Guyonnet}
\email{Jill.Guyonnet@unige.ch}
\altaffiliation{These authors contributed in equal measure to the experimental work and writing of this paper.}
\affiliation{DPMC, University of Geneva, 24 Quai Ernest-Ansermet, 1211 Geneva 4, Switzerland}
\author{H. B\'ea}
\altaffiliation{These authors contributed in equal measure to the experimental work and writing of this paper.}
\altaffiliation{now at SPINTEC, UMR 8191 CEA/CNRS/UJF, CEA Grenoble-INAC, 17 Av. des Martyrs, 38054 Grenoble Cedex 9, France  }
\affiliation{DPMC, University of Geneva, 24 Quai Ernest-Ansermet, 1211 Geneva 4, Switzerland}
\author{P. Paruch}
\affiliation{DPMC, University of Geneva, 24 Quai Ernest-Ansermet, 1211 Geneva 4, Switzerland}
\date{\today}
\begin{abstract}
In purely $c$-axis oriented PbZr$_{0.2}$Ti$_{0.8}$O$_3$ ferroelectric thin films, a lateral piezoresponse force microscopy signal is observed at the
position of 180\de domain walls, where the out-of-plane oriented polarization is reversed. Using electric force microscopy measurements we exclude electrostatic effects as the origin of this signal. Moreover, our mechanical simulations of the tip/cantilever system show that the small tilt of the surface at the domain wall below the tip does not satisfactorily explain the observed signal either. We thus attribute this lateral piezoresponse at domain walls to their sideways motion (shear) under the applied electric field. From simple elastic considerations and the conservation of volume of the unit cell, we would expect a similar lateral signal more generally in other ferroelectric materials, and for all types of domain walls in which the out-of-plane component of the polarization is reversed through the domain wall. We show that in  BiFeO$_3$ thin films, with 180, 109 and 71\de domain walls, this is indeed the case.
\end{abstract}
\maketitle

\section{Introduction}
Materials in which strong electronic correlations give rise to novel properties as a result of the coexistence, and in some cases, coupling of different orders are currently among the most intensely studied, both for their fundamental interest and in view of their potential use in multifunctional applications. Motivated in part by their recent availability as quasi-monocrystalline, epitaxially-grown thin films, much attention has been focused on multiferroic materials, combining two or more of the ferroic orders: ferroelasticity, ferroelectricity and ferromagnetism \cite{Eerenstein06}. In particular, magnetoelectrically coupling ferroelectric polarization (with its associated pyroelectric and piezoelectric properties) with magnetic ordering greatly increases the potential for multifunctional applications, and provides a rich field for fundamental study of the complex response of such materials to a broad range of external fields.

In all ferroic materials, the presence of symmetry-equivalent multiple ground states leads to the coexistence of regions or domains with different values of the order parameter, which are separated by thin interfaces or domain walls. These domain walls can themselves provide additional pathways to multifunctionality, beyond that of their parent material. For example, recent work on the multiferroic BiFeO$_3$, which couples ferroelectric and antiferromagnetic order, has shown conductivity in two of the three different ferroelectric domain wall types present in this insulating material \cite{seidel_DW_conduction_BFO}, while in SrTiO$_3$ ferroelastic domain walls have been shown to present a polar, possibly even ferroelectric character under strain gradients \cite{zubko_STO_ferroelastic_DW}. Theoretically, it has been suggested that different symmetries possible within the domain walls themselves, and also specifically the breaking of bulk symmetry at ferroelectric domain walls \cite{morozovska_resolutionfunction_PFM} could lead to unusual behavior.

The presence of such additional functionalities at domain walls provides an exciting new opportunity to study the interplay between different ferroic orders, symmetry and strain. In addition, the intrinsically nanoscale size of the domain walls makes them interesting as potential device elements for ever smaller applications. However, it is this very small size that also presents the greatest challenge to experimental investigations of the behavior and properties of individual domain walls.

One of the primary techniques developed in recent years, allowing the study of individual nanoscale systems at the requisite size scales, has been scanned probe microscopy, in which different interactions between a specially fabricated probe and the sample surface can be measured with (sub)nanometric precision. For insulating materials such as ferroelectrics and multiferroic materials, various applications of atomic force microscopy (AFM), such as conductive-AFM and in particular piezoresponse force microscopy (PFM), used to study ferroelectric polarization \cite{guthner_dransfeld_localpoling,gruverman_sfm1996,nemanich_pfm1998}, have been especially useful.

In PFM, a metallic AFM tip is used to apply a periodic voltage across the ferroelectric material, locally exciting a complex mechanical response at the film surface. This piezoelectric response is in turn detected as the first harmonic component of the deflection of the AFM probe cantilever, recorded by the position of a laser beam reflected from the cantilever onto a quadrant-split photodiode. Detection of the vertical cantilever deflection and its angular torsion are referred to as vertical and lateral PFM, respectively (VPFM and LPFM). In each case, the response phase provides information on the polarization state, while its amplitude is related to the polarization magnitude. Depending on the piezoelectric coefficient tensor $d_{ij}$, linked to the crystal symmetry, a combination of these two measurements allows access to both out-of-plane and in-plane components of the polarization (see review by Kalinin {\it et al.} \cite{kalinin_PFM_review} and references therein).  The situation becomes much more complex when structural variations of the film surface are encountered \cite{peter_PFM_shapeeffects}, or when the sample presents a granular morphology of unknown crystalline orientation \cite{eng_3D_PFM}. In addition, artefacts due to cross-coupling between mechanical and electrostatic effects may be present \cite{jungk_vectorPFM_challenge,jungk_background_PFM}, and have to be taken into account. However, in spite of these complications, PFM has yielded a host of data on ferroelectric domain wall behavior, including detailed examinations of their static configuration and dynamic response, governed by the competition between elasticity and disorder \cite{tybell_creep,paruch_dynamics_FE,paruch_DW_roughness_FE}, domain wall interaction with a strong localized defect or tip potential \cite{morozovska_DW_defect,morozovska_DW_tip}, and the possible influence of magnetoelectric coupling in BiFeO$_3$ \cite{catalan_BFO_DW}. Applying PFM specifically to the question of symmetry and symmetry-breaking within domain walls would be a natural and powerful extension of this technique.

These effects of symmetry breaking or change may be observed even in the relatively ``simple'' case of a single-crystal-like $c$-axis-oriented tetragonal ferroelectric film (ie. with a purely out-of-plane polarization, directed along the $c$-axis of the film). From symmetry considerations, the  $d_{35}$ and $d_{34}$ piezoelectric coefficients are zero, and thus no lateral piezoresponse is expected under the application of an out-of-plane electric field. However, recent numerical analyses in the framework of resolution-function theory show that  at the position of 180\de domain walls in such a film, a local shear originating from strain effects \cite{morozovska_resolutionfunction_PFM} leads to an horizontal displacement of the surface at the domain wall. This displacement has a similar effect to that of a non-zero $d_{35}$ (or $d_{34}$) bulk crystal coefficient, giving rise to a lateral PFM signal. For clarity we label this local enhacement of domain wall shear $d_{35}^{DW}$  (similarly to the $d_{35}^{eff}$ notation of reference \cite{morozovska_resolutionfunction_PFM}).  Although such signals had already been observed in tetragonal materials like PbZr$_{0.2}$Ti$_{0.8}$O$_3$ (PZT), they were previously attributed to other mechanisms, including purely vertical height differences between the regions with opposite $c$-axis polarization \cite{ganpule_3D_PFM} resulting in tilting or torsion of the AFM cantilever \cite{wittborn_domain_imaging,scrymgeour_pfm_180DW}, or to surface electrostatic effects at domain walls \cite{jungk_DW_LPFM}. We therefore decided to carry out a detailed PFM investigation of such domain walls to determine the microscopic origin of the observed LPFM response

In this article, we report on the LPFM signal observed at 180\de domain walls in tetragonal ferroelectrics, which we attribute to shear strain resulting from the symmetry-breaking sign change of the vertical deformation (linked to the $d_{33}$ coefficient) across the domain wall. After a review of the materials and experimental techniques used in our studies in section II, we present the experimental observations of shear strain leading to LPFM response at PZT domain walls in section III. By comparing the time evolution of the PFM signal, which remains undiminished over the month-long duration of our experiments with electric force microscopy measurements, whose amplitude decreases over time, we show in section IV that the observed signal is not of electrostatic origin. With finite element simulation of the tip-sample interaction during PFM, and topographical analysis of the domain walls under a constant applied bias, in section V, we also exclude surface tilting as a potential microscopic mechanism behind the LPFM response. The details of the shear-strain scenario, including simulations of the resulting cantilever deflection and torsion, are discussed in section VI. Finally, in section VII, we consider the LPFM response at different types of domain walls in BFO thin films grown in different crystallographic orientations, showing how the response to both shear strain and in-plane components of polarization contributes to the final LPFM signal.

\section{Methods}
\label{section_methods}
\subsection{Growth}
\begin{figure}[ht]
\includegraphics[width=0.7\columnwidth]{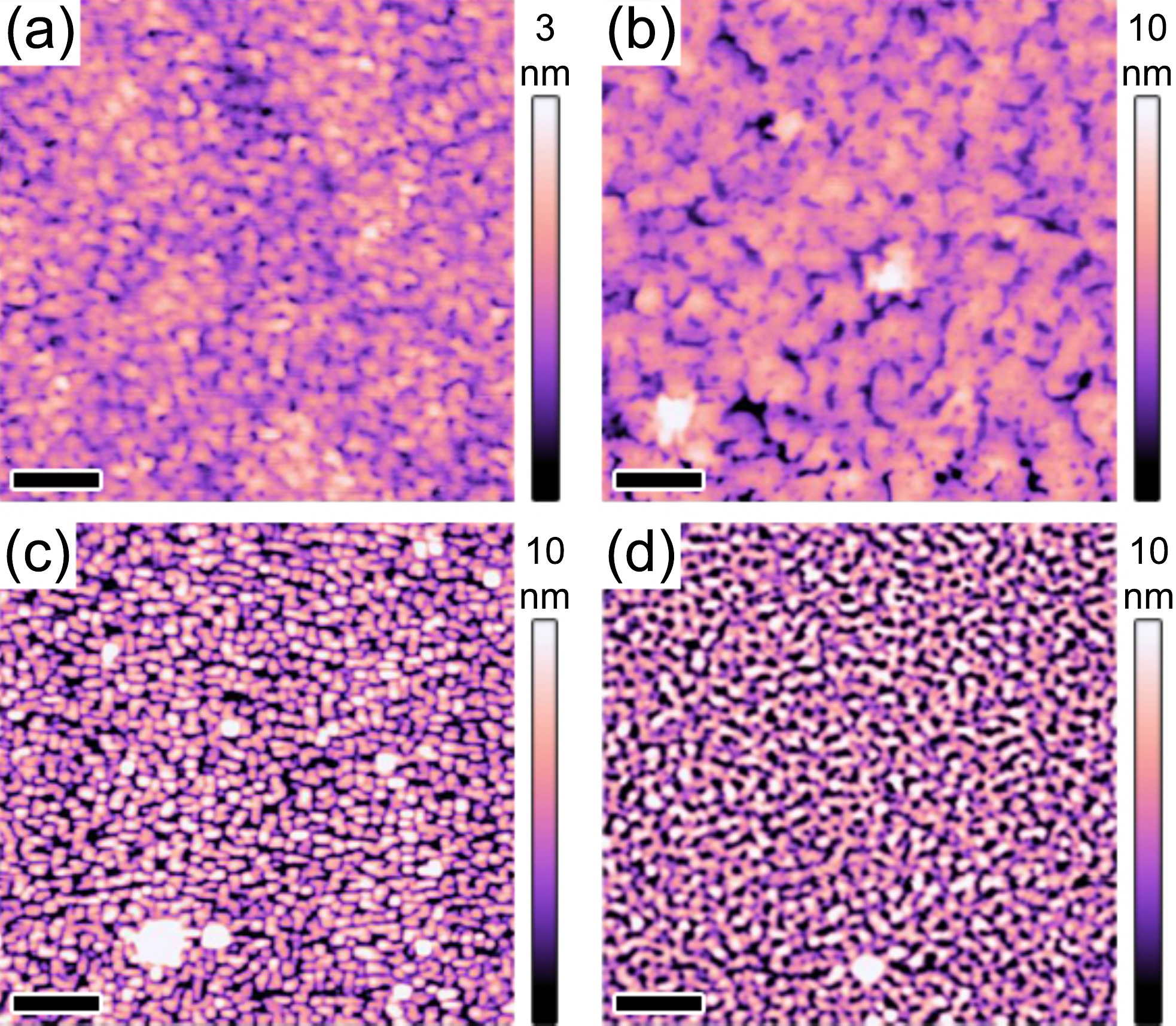}
\caption{Surface topography of 70 nm $c$-axis PZT (a), 250 nm PZT including $a$-axis domains (b), (001)-BFO (c) and (111)-BFO (d) thin films performed by AFM, showing rms roughnesses of 4 $\mbox{\AA}$ for $c$-axis PZT, 1.3 nm for thicker PZT and 2 nm for BFO. Horizontal scale bar is 0.5 $\mu$m in (a) and (b), 0.6 $\mu$m in (c) and 0.7 $\mu$m in (d).}
\label{fig_roughness}
\end{figure}

PZT and BFO films were epitaxially grown on SrTiO$_3$ substrates by sputtering and pulsed laser deposition, respectively. AFM topography studies (\fref{fig_roughness}) show their typical surface rms roughness to be 0.4 nm for 70nm PZT, 1.3 nm for 250 nm PZT and 2 nm for BFO. Bottom
electrodes of SrRuO$_3$ or La$_{0.7}$Sr$_{0.3}$MnO$_3$ (in the case of (111)-BFO only) were used. As demonstrated by x-ray diffraction studies, the PZT films are tetragonal, with a monodomain ``up'' polarization as grown. The thinner films (70 nm) are purely $c$-oriented, with ``cube-on-cube'' growth on the (001)-STO substrates, while thicker films (250 nm) also show the presence of a small amount of $a$-domains \footnote{At the resulting 90\de domain walls no reversal in the $d_{33}$ coefficient sign occurs, and the resulting LPFM signal is complex, and rather difficult to interpret quantitatively given the narrowness of the a-domains. These domain  walls will thus not be discussed in the paper.}. In the case of (111)-BFO, the structure is rhombohedral with the trigonal axis (and thus the polarization) oriented along the growth axis, out-of-plane of the film \cite{BeaAPL111}. For (001)-BFO, although initial x-ray measurements indicated a tetragonal symmetry \cite{Beaphilmag} local measurements of the polarization (as discussed in section VII) show both in- and out-of-plane components, thus suggesting a monoclinic structure in which the polarization lies close to the $<$111$>$ crystallographic axes. At the coherence length of the x-rays used for the measurements, the small size of the intrinsic domains leads to averaging over the 8 different polarization variants, thus presenting an effective tetragonal structure.
\subsection{Piezoresponse Force Microscopy measurements}
AFM measurements were made at ambient conditions with a \emph{Nanoscope Dimension V} with \emph{NCS18 Cr-Au} tips from \emph{MikroMasch} or \emph{MESP} from \emph{Veeco} presenting similar geometrical characteristics \footnote{Cantilever length: 230 $\mu$m; cantilever thickness: 4 $\mu$m; cantilever width: 40$\mu$m; tip length: 20 $\mu$m; opening angle 40\de}. PFM images were performed in contact mode at 20 kHz, with ac voltage amplitude of 2.5-5 V, and a deflection setpoint of 0.5 V. We also performed double-pass PFM, interleaving two measurements on each scan line, with the first pass in the strong-contact regime just described, and the second pass made with different deflection setpoints, and thus with different forces between the tip and the substrate, so as to be in either weak-contact or even in non-contact (tip/surface separation around 5 nm) regimes. In such measurements of the weak and non-contact regimes, the lower mechanical coupling between the tip and the surface \cite{kalinin_PFM_FE} would give a greater weight to possible electrostatic contributions, allowing us to explore their role in the origin of the LPFM signal at domain walls.
\subsection{Electric Force Microscopy measurements}
To further measure the purely electrostatic interaction between the AFM tip and the sample surface, we performed electric force microscopy (EFM) measurements, interleaving standard topographic tapping mode line scans with lift mode, in which the tip is scanned at a constant distance above the surface, and electrostatic interactions can be measured through the phase shift of the tip resonance. In order to minimize the contributions from topographical and short range van der Waals forces, relatively large distances were chosen. Using either a tapping-mode-like excitation to obtain vertical oscillation of the cantilever (vertical EFM or VEFM) or a torsion mode excitation to obtain an oscillation of the tip around the cantilever axis (torsional EFM or TEFM), we were able to access both out-of-plane and in- plane components of the electric field. In each case, the presence of an electrostatic force along the direction of the tip oscillation near resonance (due to a vertical electric field for VEFM, and a horizontal electric field, perpendicular to the cantilever axis, for TEFM) leads to a change of phase or amplitude of this response, giving rise to a signal.

In the VEFM case, the first vertical bending resonance around 70 kHz was used while for the TEFM measurements, the tip was excited in the lateral
bending resonance mode around 800 kHz. 

\subsection{Simulations}
To simulate the interaction between the tip and the ferroelectric material, and to compute  the resulting cantilever deflection and torsion in the different possible scenarios, we used \emph{COMSOL Multiphysics 3.5}, with the geometry of a typical \emph{NCS18 Cr-Au} tip. The position of the base of the cantilever, which is clamped to the cantilever holder in the actual experiment, was set at a fixed position and a force was applied at the tip apex. The applied force was a superimposition of a vertical force of 10 nN and a varying vertical or horizontal force (between 0 and 10n N) in order to simulate either a horizontal or a tilted surface below the tip and compare the results with the $d_{33}$ vertical
piezoresponse.
\section{Lateral PFM at domain walls}
In this section, we show VPFM and LPFM measurements in PZT ``up''-polarized monodomain films on which we have written ``down'' domains by applying a positive voltage while scanning the tip over the surface of the designated area.
\begin{figure}[ht]
\includegraphics[width=\columnwidth]{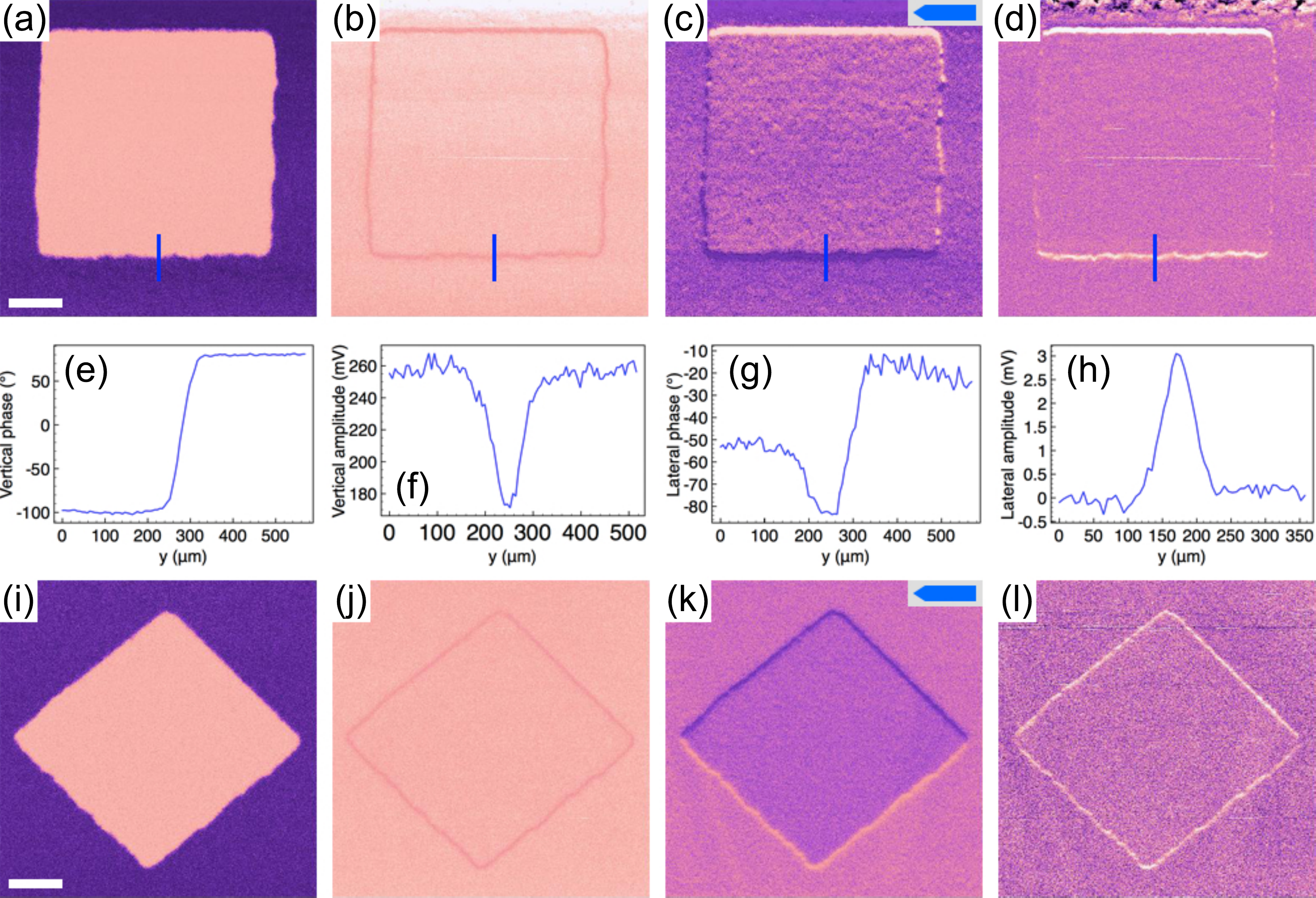}
\caption{PFM measurements of a square ``down''-polarized ferroelectric domain
written on an ``up''-polarized 70 nm PZT film by applying +8V while scanning
over a 2$\times$2$\mu$m$^2$ area. The VPFM measurement shows a 180\de phase
contrast [(a) and (i)] with a corresponding minimum in the amplitude at the domain wall [(b) and (j]. In the LPFM signal, two opposite features are observed in the phase at the domain walls [(c) and (k)], perpendicularly to the AFM cantilever, with a corresponding rise in the amplitude [(d) and (l)]. The orientation of the cantilever is indicated in inset in (c) and (k). Horizontal scale bar is 0.5 $\mu$m for (a)-(d) and 0.6 $\mu$m for (e)-(h). The fast scan direction, the same for all measurements, is along the cantilever length. For the first measurement set, line cuts corresponding to the vertical blue lines in [(a)-(d)] are shown in [(e)-(h)].}
\label{fig_pztpfm}
\end{figure}

In a measurement performed immediately after writing with +8V over a 2$\times$2$\mu$m$^2$ area, we observe \cite{guyonnet_lpfm_apl} the expected 180\de contrast in the VPFM phase between ``up'' and ``down''-polarized regions (\fref{fig_pztpfm}(a)). Correspondingly, a minimum in the VPFM amplitude is seen at the domain walls (\fref{fig_pztpfm}(b)). In the simultaneously recorded LPFM measurement, a nonzero phase signal is observed only at the position of the domain walls which are parallel to the cantilever axis (\fref{fig_pztpfm}(c)), as confirmed by maxima in the amplitude. Moreover, the sign of the phase signal depends on the relative positions of the ``up'' and ``down'' domains with respect to the AFM cantilever axis when it is at the position of the domain wall  \footnote{We note here that the LPFM phase reference is frequency-dependent, as reported previously by Jesse \textit{et al.} \cite{jesse_dynamic_pfm}. Moreover, we noticed that the phase reference may change when the lock-in is switched on and off or when the frequency or ac amplitude is changed. Therefore it may not yield a reproducible contrast phase from one measurement to another. This is the reason why the contrast is inverted between \fref{fig_pztpfm}(c) and \fref{fig_pztpfm}(g). However, we have checked with other setups that this phase change between two measurements at the same frequency has no physical origin and is linked to the internal lock-in phase reference change.}. This observation was confirmed when we manually rotated the sample by 45\de to change the relative position of the cantilever with respect to the written domain (\fref{fig_pztpfm}(i)-(l)). The domain walls which are perpendicular to the cantilever in \fref{fig_pztpfm}(c)-(d), for which the response would be along the cantilever axis and thus non-detectable by the torsional response around this axis, do not give a strong LPFM signal. When the sample is rotated by 45\de (\fref{fig_pztpfm}(k)-(l)), these same walls, for which the lateral signal now has a non zero projection perpendicular to the cantilever, do give a strong LPFM signal.

Such an LPFM signal had been observed in previous studies \cite{ganpule_3D_PFM,gruverman_caps_afm,gruverman_tutorial_aveiro} and
attributed to a number of different mechanisms. Initially, a surface deformation due to opposing vertical contraction and expansion on either side of the domain wall, inducing a sliding of the tip on the tilted surface was suggested by Wittborn {\it et al}. \cite{wittborn_domain_imaging}. The same qualitative mechanism was later investigated by Scrymgeour {\it et al}., who assumed that the tip friction could lead to a torsion of the cantilever \cite{scrymgeour_pfm_180DW}. Both explanations were however contested by Jungk {\it et al}. \cite{jungk_DW_LPFM}, who in turn proposed the contribution of the electrostatic interaction between the tip and the electric field arising from the change in sign of (unscreened) charges at the ferroelectric surface at the domain wall. More recently, numerical analyses in the framework of resolution-function theory by Morozovska {\it et al}.  \cite{morozovska_resolutionfunction_PFM} showed that in a $c$-axis tetragonal film, where the $d_{35}$ and $d_{34}$ piezoelectric coefficients are zero by symmetry in uniformly polarized regions, the local breaking of symmetry at the 180\de domain wall induces shear that leads to an effect similar to the one of finite $d_{35}$ and $d_{34}$ values (thus giving an effective $d_{35}^{eff}$ coefficient), potentially giving rise to an LPFM signal \footnote{In fact, depending on the direction of the domain wall, the signal will result from a linear combination of the  $d_{35}$ and $d_{34}$ coefficients, equivalent in a tetragonal material like PZT. In the following, we will only use the $d_{35}$ term to account for the resulting piezoelectric coefficient.}. 

\section{Contribution from electrostatic forces}
In PFM measurements, great care must be taken to correctly account for possible electrostatic contributions to the signal, which may be of the same order of magnitude as the mechanical response due to the local piezoelectric excitation of the material, depending on the specific contact force and tip parameters used \cite{kalinin_PFM_FE}.  A strong electrostatic interaction could lead to artefacts such as the mechanism proposed by Jungk {\it et al}. \cite{jungk_DW_LPFM} to explain the observed LPFM signal at domain walls. To investigate the extent of the electrostatic contribution to our PFM signal, and particularly to the lateral signal at domain walls, we performed double-pass PFM measurements of linear domains (written with $\pm$ 10 V in a 250 nm thick PZT film) with different contact forces down to zero force, i.e. in a non-contact configuration.
\begin{figure}[ht]
\includegraphics[width=\columnwidth]{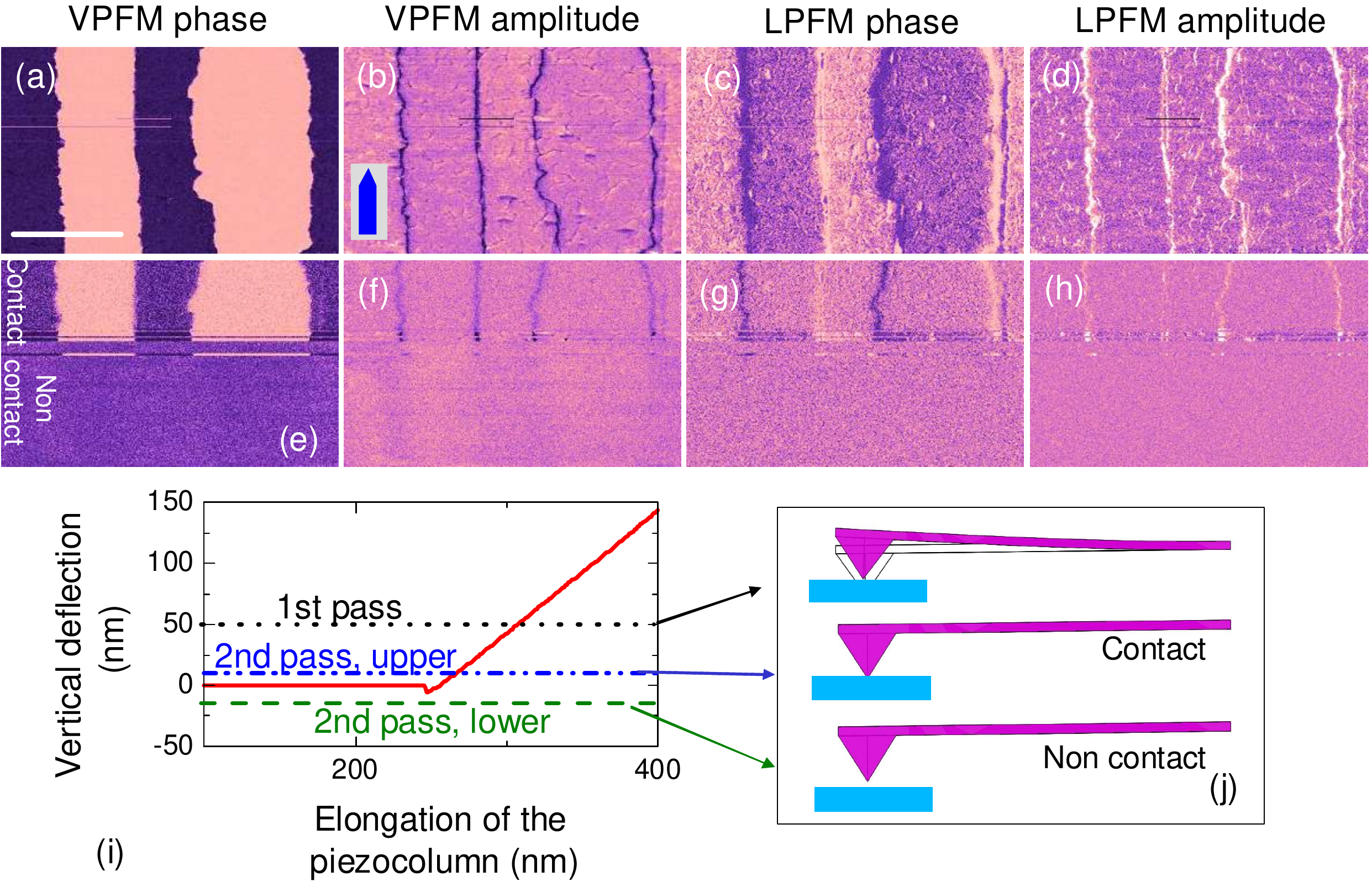}
\caption{Double-pass PFM measurement of domains written on a 250 nm PZT film by applying +10/-10/+10V to the tip. The first pass [(a)-(d)] is done in strong contact with a setpoint of 0.5 V, while the interleave pass [(e)-(h)] is performed with different lift heights providing weak contact in the upper part and no contact in the lower part, as indicated in (e). (i) Typical force curve (in red solid line) with the strong, weak and non-contact configuration in the first and second passes (upper and lower part of (e)-(h), see horizontal dotted or dashed lines giving the used setpoint for the different passes). The cantilever orientation is given for all measurements in (b). The fast scan direction is perpendicular to the length of the cantilever. The scale bar is 1$\mu$m. (j) Representation of the tip/cantilever and sample configurations in the three regimes described in (i).}
\label{fig_pfmint}
\end{figure}

The first pass was done in contact with a deflection setpoint of 0.5 V, as represented by the strong contact regime in \fref{fig_pfmint}(i) and (j). As previously, the VPFM measurement shows a 180\de contrast in the phase response corresponding to ``up'' and ``down'' polarized regions (\fref{fig_pfmint}a), with a decrease in amplitude at the position of the domain walls (\fref{fig_pfmint}b), while a dark (resp. bright) contrast on the left (resp. right) of the written domain can be seen in the LPFM phase signal at domain walls (\fref{fig_pfmint}c), associated with a strong increase in the LPFM amplitude (\fref{fig_pfmint}d).

Following the topography acquired during this first pass, but this time with a different applied force, a second PFM pass was performed for every scan line. Initially (upper part of \fref{fig_pfmint}(e)-(h)), a small force was applied, resulting in weak contact between the tip and the surface, as represented by the weak contact regime in \fref{fig_pfmint}(i) and (j). Partway through the measurement (lower part of \fref{fig_pfmint}(e)-(h)), the tip-surface force was decreased even further by lifting the tip above the surface (non-contact) \footnote{In the region separating these two configurations, the setpoint was increased to obtained a non contact regime. Instead, we obtained an unstable oscillation between the weak and non-contact regimes, and thus had to further decrease the deflection setpoint to ensure a stable non-contact regime.}. In the weak contact regime, the obtained results for both VPFM and LPFM signals are similar to those measured during the strong contact first pass, although more noisy. In addition, the observed LPFM phase signal at the domain wall is narrower in the second pass. In the non-contact configuration, the contrast disappears in both VPFM and LPFM phase reponses (\fref{fig_pfmint}(e) and (g)), as does the increased LPFM amplitude at domain walls \fref{fig_pfmint}(h). The decrease in VPFM amplitude associated with domain walls also disappears, although we note a faint contrast between the two different polarization orientations (\fref{fig_pfmint}(f)), strongly reminiscent of an electrostatic signal, and possibly due to the electrostatic interaction in the non-contact regime. 

From these observations we can conclude that mechanical rather than electrostatic interactions are the dominant contribution to the observed LPFM response, since it is when the {\it mechanical} contributions are minimized by lifting the tip into a non-contact regime above the surface that the LPFM signal disappears. Indeed, the stronger signal during the first pass is consistent with the better mechanical coupling in a strong contact regime, allowing the tip to more easily follow the lateral deformation of the surface at the domain wall. As the mechanical coupling worsens with decreased tip-surface contact force, the tips follows this lateral deformation less easily, resulting in narrower observed features, and generally a lower signal to noise ratio. We emphasize here that similar results are obtained for thinner PZT and BFO films.

To further explore the extent of electrostatic contributions at domain walls, we performed EFM measurements on similar domains written in a 70 nm PZT sample. We expect these measurements to be even more sensitive to electric forces than the double-pass PFM contact mode measurements, since EFM images are acquired in a non-contact mode, with the cantilever close to resonance.

\begin{figure}[ht]
\includegraphics[width=\columnwidth]{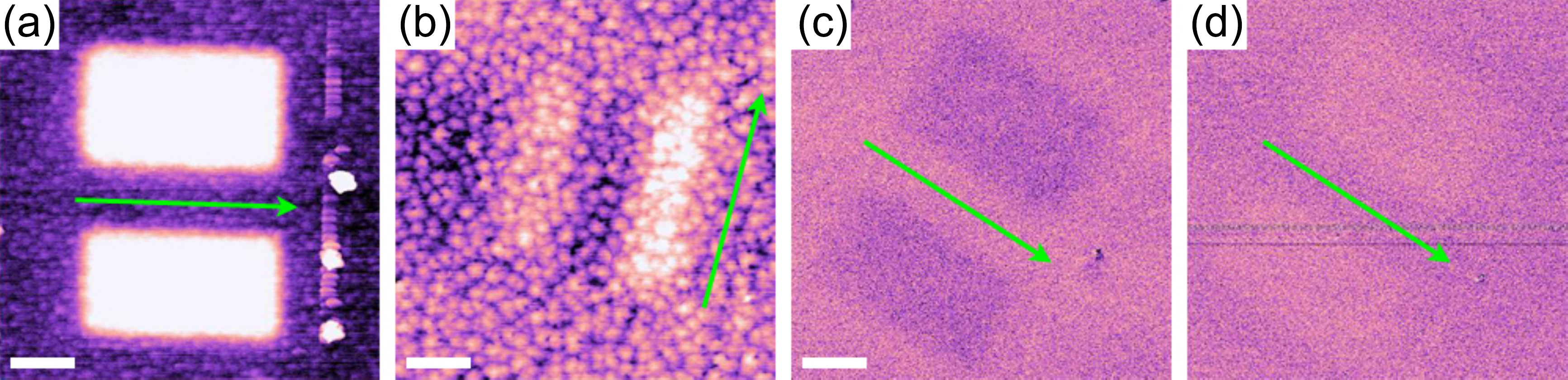}
\caption{EFM measurements performed in tapping [(a) and (b)] and torsion [(c)
and (d)] modes on domains written on a 70 nm PZT film. The VEFM signal measured
in the same set of domains immediately after domain writing (a) and one month later (b) shows an intensity decrease of an order of magnitude (both measurements were performed with a tip
lift of 30 nm and a +1V dc bias applied to the tip and are shown with a vertical scale of 2\de). The TEFM signal observed with +3V (c) and -3V (d) dc bias applied to the tip immediately after writing was not observable after one month (not shown). The white bar is  1.2 um (a),  0.55 um (b) and 1.1 um (c)-(d). The orientation of the long axis of the rectangular domains is indicated by the green arrow in each measurement.}
\label{fig_pztefm}
\end{figure}
The strong VEFM contrast observed immediately after domain writing [\fref{fig_pztefm}(a)], diminished progressively with time, decreasing by an order of magnitude over one month [\fref{fig_pztefm}(b)], all measurements being performed in the same conditions (30 nm tip lift and +1V dc bias applied to the tip). However, as we previously reported \cite{guyonnet_lpfm_apl}, both
VPFM and LPFM signals remained of comparable magnitude throughout the duration of the experiment. The different time evolution of the electrostatic and piezoelectric signal therefore strongly suggests that the observed LPFM response at domain walls is largely independent of electrostatic interactions.

To directly investigate the presence of closure electric fields at domain walls as proposed by  Jungk {\it et al}. \cite{jungk_DW_LPFM}, we also carried out TEFM measurements to check for the possible presence of a strong horizontal electric field at the domain wall expected in this scenario, resulting from unscreened positive charges on one side of the domain wall and negative charges on the other side.  In this mode, the cantilever is excited at a lateral torsion resonance. Thus, the presence of a horizontal electric field perpendicular to the cantilever axis will lead to a horizontal force acting on the charged tip, and should change the resonance phase, giving rise to a TEFM signal. We note that the electric field at the origin of the mechanism proposed by Jungk {\it et al}. for LPFM signal at domain walls is assumed to occur to counteract the depolarizing field present in an unscreened configuration of alternately polarized domains.  More generally, such a scenario could also originate from overscreening of the ferroelectric charges during writing with the AFM tip, and would in this case be in the opposite direction. In either case, the presence of such horizontal electric fields would lead to both a VEFM signal with positive or negative contrast on the domain, and a concomitant TEFM signal at the domain walls. 

However, as show in \fref{fig_pztefm}(c) and (d) TEFM images taken in the same region as in \fref{fig_pztefm}(a) immediately after domain writing, with a tip lift of 15 nm and $\pm$3V applied to the tip, only a very faint contrast (0.1\de) is observed between ``up'' and ``down'' polarized regions, its sign depending on the sign of the tip bias, similarly to the VEFM images. Such coupling between VEFM and TEFM could be due to a small vertical oscillation of the tip associated with the cantilever torsion.  More importantly, no particular features are observed at the domain walls, as would be expected in the presence of the proposed horizontal field. Moreover, we observed that this weak TEFM signal had completely disappeared during the next measurement, one month later.

The decrease of the EFM signal over time can be understood as the passivation of the excess screening charges brought to the surface during domain writing, and which lead to a strong initial EFM response.  These excess charges are then passivated to equilibrium \footnote{The passivation of the surface charge depends both on the quality of the ferroelectric film, since the presence of defects can give rise to mobile charge carriers in these nominally insulating materials, and on the external environment, which can provide surface screening charges.} and the EFM signal decreases accordingly. In the specific measurement of \fref{fig_pztefm}, 2 ``down''-polarized domains were written with +10 V, separated by an unwritten region with the as-grown monodomain ``up'' polarization, giving rise to a repulsive interaction (bright contrast in VEFM) with a positively biased tip, separated by the same mid-range contrast as that observed over the rest of the unwritten background.  With time, this bright contrast becomes closer and closer to the same mid-range level.

In summary, we have shown in this section that electrostatic interactions contribute only weakly to the LPFM signal observed at 180\de domain walls in PZT, and may thus be neglected in the search for the microscopic mechanisms leading to this phenomenon.

\section{Contributions from vertical surface deformations and roughness}
Having shown that mechanical contributions dominate in the observed LPFM signal at 180\de  domain walls, we now discriminate between different possible scenarios. As a preliminary observation, it has been shown that the surface morphology of the sample can lead to significant artefacts, especially in the LPFM signal \cite{peter_PFM_shapeeffects}. However, the very low surface roughness of the PZT films, along with the fact that the observed features can be seen only at domain walls, allow this hypothesis to be eliminated. The observed LFPM response is  clearly associated with domain walls themselves, and is not an effect of features on the film surface.

\begin{figure}[ht]
\includegraphics[width=0.8\columnwidth]{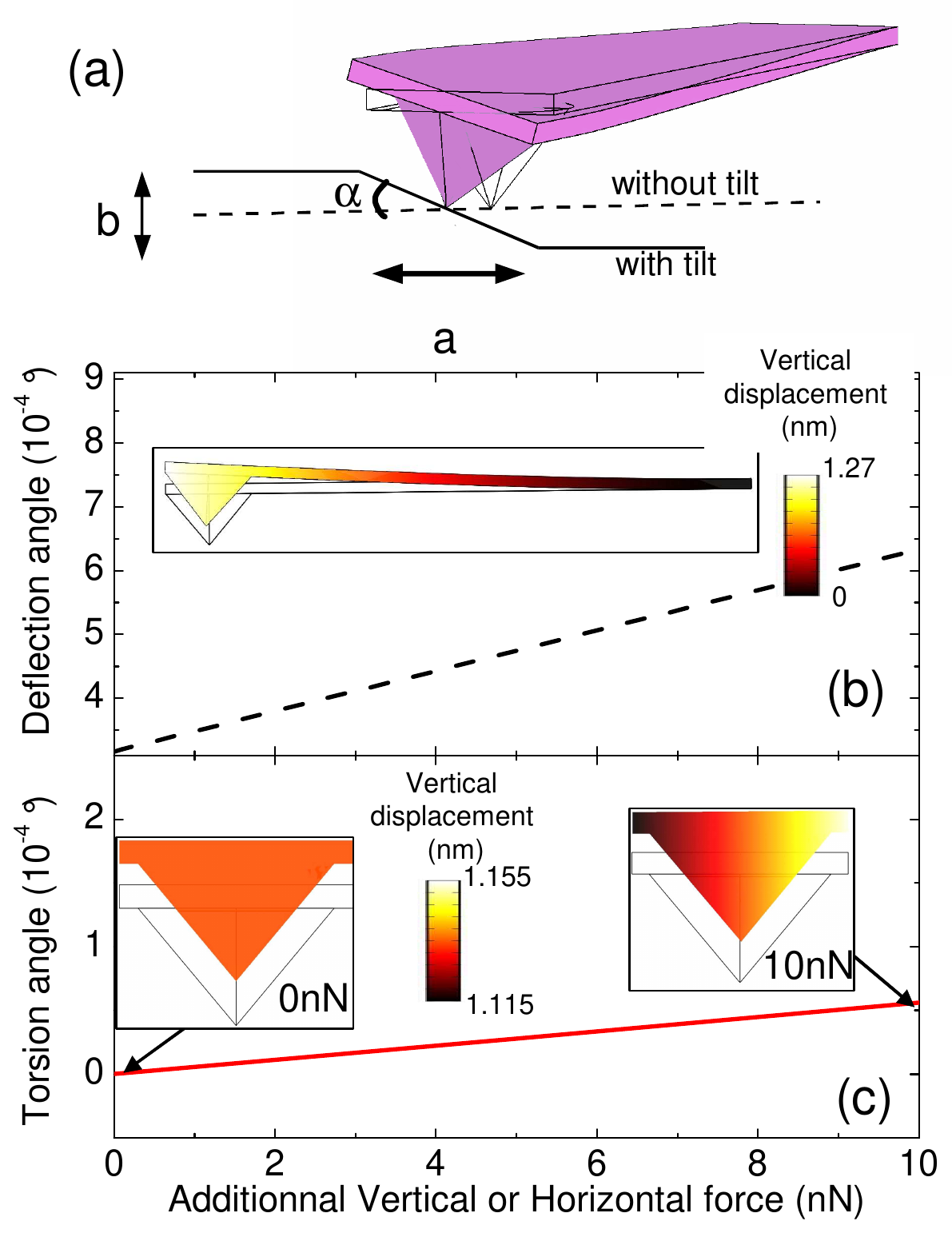}
\caption{(a) Representation of the torsion of a cantilever/tip system that could occur if the underlying surface is tilted by an angle $\alpha$ at a domain wall, in block color. The state without surface tilting is represented by the dotted line, and the line-sketch of the cantilever.  Evolution of vertical deflection ((b), dashed line) and horizontal torsion ((c), red line) angles as a function of the applied vertical (b) or horizontal (c) force with a superimposed constant vertical force of 10nN applied at the tip apex. The inset in (b) shows the side view of the vertical deformation of the cantilever. The two insets in (c) represent front views of the deflection and torsion of the cantilever/tip system for 0 and 10 nN horizontal forces, as indicated by the arrows. In all the insets, the initial tip and cantilever position are indicated by the line sketches, while the colors correspond to the vertical displacement given in nm in the color scales for side and front views.}
\label{fig_horvervague}
\end{figure}

We then investigated the scenario proposed by Wittborn {\it et al}. \cite{wittborn_domain_imaging}, where the opposite vertical deformation on either side of the domain wall results in a tilt of the surface below the tip, and can lead to torsion of the cantilever. In order to quantify this contribution, we first calculated the tilt angle of the surface at a 180\de domain wall in PZT, as illustrated by \fref{fig_horvervague}a. Under the action of an out-of-plane electric field $E_z$ corresponding to an applied voltage $V$ of 1V across the thin film, we expect the elongation on one side of the domain wall and contraction on the other due to the inverse piezoelectric effect, thus giving a total height difference at the surface of $2d_{33}\times V$. The surface, being continuous, locally tilts at the domain wall, with the lateral extension of this tilted region $\Delta x$ depending on the anisotropy and elastic constants of the material. For PbTiO$_3$ ($d_{33}$=75pm/V), Morozovska {\it et al}. \cite{morozovska_resolutionfunction_PFM} have calculated that the change of height occurs over a distance of $\Delta x=40$ nm. The tilt angle $\alpha$ of the surface at a domain wall is thus given by:
\begin{eqnarray}
\begin{split}
\alpha&=arctan(\frac{b}{a})=arctan(\frac{2d_{33}\times V}{\Delta x})\\
&=arctan(\frac{150.10^{-12}}{40.10^{-9}})=0.21\,^{\circ}
\end{split}
\label{alpha}
\end{eqnarray}

Using \emph{COMSOL Multiphysics}, we then modeled a tip in contact with a horizontal or tilted sample surface, setting the cantilever base at a fixed position while applying a force at the tip apex, as detailed in section \ref{section_methods}. Calculations were carried out with a constant vertical force of 10 nN, corresponding to the contact force, resulting in a vertical displacement of the cantilever with an angle $\beta$=$3.1\cdot 10^{-4}$$^{\circ}$ (see definition of $\beta$ in \fref{fig_anglestip}), as represented in inset of  \fref{fig_horvervague}(b). Onto this vertical force component, we superimposed varying vertical or horizontal forces of up to 10 nN that would account for vertical piezoelectric response and tilt of the surface, respectively. In this scenario, a slight tilt of the surface would also tilt the surface normal away from its initially vertical position, thus inducing a small horizontal component in the contact force. A zero horizontal force would thus
correspond to an untilted horizontal surface and a 10 nN horizontal force to a surface tilted by 45\de, that, although unphysically large, was considered for completeness. Representations of the distortion of the cantilever/tip system are given in insets of \fref{fig_horvervague}(b) (side view) for purely vertical deflection and \fref{fig_horvervague}(c) (front view) for different horizontal forces.

From \fref{fig_horvervague}(b) and (c), we can see that both deflection and torsion angles vary linearly when the corresponding force increases. However, the variation of the torsion angle with a horizontal force is much lower than that of the deflection angle with a vertical force, as indicated by the different gradients of the corresponding graphs.  Thus, the tip remains on the surface when deflected by a given vertical force, but it will not remain perpendicular to the surface and will not follow the surface tilt. In this range of vertical and horizontal forces, the variation of deflection angle is around $3.1\cdot 10^{-5}$$^{\circ}$ per nN and the variation of torsion angle is one order of magnitude below, around $5.6\cdot 10^{-6}$$^{\circ}$ per nN. This is evidenced in the insets of \fref{fig_horvervague}(b) and (c). For a purely vertical force, i.e. an untilted surface, only a vertical deflection of the cantilever occurs, as shown in the inset of \fref{fig_horvervague}(b), and no torsion is obtained, demonstrated by the uniform mid-scale color of the lower left inset of \fref{fig_horvervague}(c). For a horizontal force of 10nN, corresponding to an unphysically large tilt of the surface, we begin to observe a small torsion of the cantilever superimposed on the vertical deflection, as shown by a non-uniform vertical displacement of the cantilever/tip system (see right inset of \fref{fig_horvervague}(c)).

From these calculations, we extract values for the deflection and torsion angles that would give rise to vertical and horizontal signals on the photodiodes, respectively, as defined in \fref{fig_anglestip}.
A typical vertical piezoelectric response would lead to a vertical displacement of $d$=150pm. By taking a force constant $k$=3N/m, we obtain an estimate for the vertical force due to piezoelectric response of $F=kd=0.45$nN (that has to be added to the contact force).  This would correspond to a deflection angle of $1.4\cdot 10^{-5}$$^{\circ}$. From the tilt of the surface by an angle of $0.21$$^{\circ}$, as calculated in Eq. \ref{alpha}, the horizontal force would be $F_{horizontal}=\sin(0.21^{\circ})=3.7\cdot10^{-3}\times F_{vertical}$. The corresponding torsion angle would thus be $5.6\cdot10^{-6} \times 3.7\cdot10^{-3} \times F_{vertical}=2.1\cdot10^{-9}\times F_{vertical}$.

In our measurements, both vertical and horizontal signals have similar magnitudes \footnote{In imaging, the LPFM signal is amplified by a factor 16; this is however corrected in our simulations.}. In order to obtain a torsion angle in the same range as the deflection angle, one would thus have to have a contact force of $F_{vertical}$=$4.4 \cdot10^{3}$nN, which is well above the typical contact
forces (usually in the range between few tens and hundreds of nN with the tips used in our experiments), thus excluding the tilting of the surface at the domain wall as a realistic mechanism explaining the lateral PFM signal.

\section{Shear signal at domain walls}
\begin{figure}[ht]
\includegraphics[width=0.5\columnwidth]{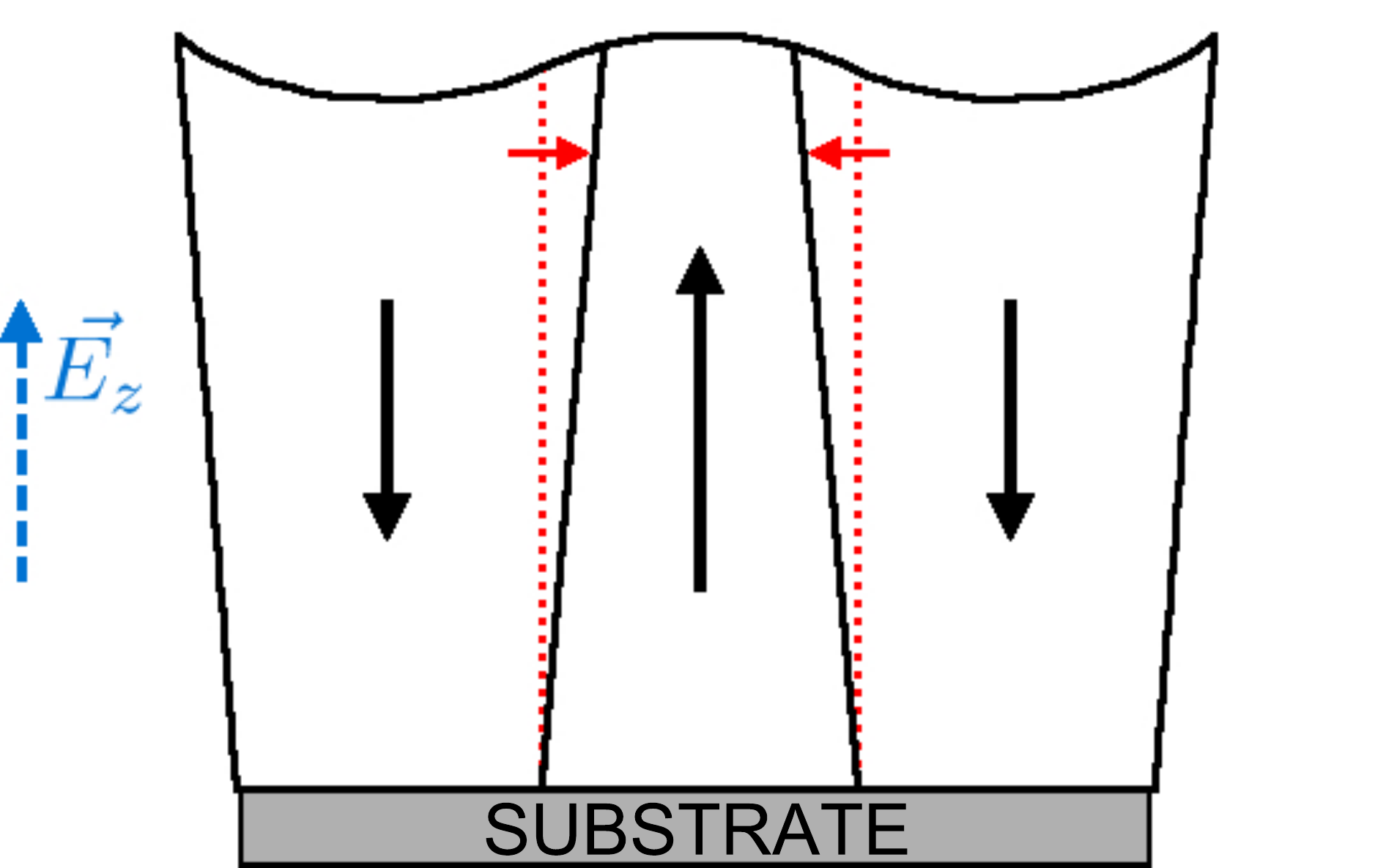}
\caption{Sketch of domain wall shear in presence of an out-of-plane field $E_z$,
indicated by the dashed arrow. The initial domain wall positions are indicated by
the dotted lines. The horizontal red arrows represent the lateral motion of the
domain walls.}
\label{fig_d35shear}
\end{figure}

Having shown that neither electrostatic effect, nor surface tilting present a satisfactory explanation of the microscopic mechanism behind the LPFM signal observed at 180\de domain walls in $c$-oriented PZT thin films, we therefore turn to the one scenario that does appear to agree with the experimental observations,  that of a shear displacement occurring specifically at domain
walls. A simple representation of the domain wall using consideration of basic elasticity and unit cell volume conservation can be used to qualitatively model the observed behaviour.  In this case, as a result of the antiparallel orientation of the polarization on the two sides of the domain wall, the vertical deformation of the material in the presence of an out-of-plane field (due to the $d_{33}$ piezoelectric coefficient) will also be of opposite sign: an ``up''-polarized domain in a positive field will expand, while its ``down''-polarized neighbor will contract. However, by the Poisson effect, this vertical deformation will also result in lateral changes (the ``up''-polarized cell contracting laterally, the ``down''-polarized cell expanding laterally as a result of a positive applied out-of-plane field, which may also be expressed by the $d_{31}$ coefficient). In a uniformly polarized bulk sample, this effect, being symmetric around the PFM tip, would not lead to a lateral response. However, at domain walls in a ferroelectric thin film, clamped to the non-ferroelectric substrate on which it is epitaxially grown, the upper region of the domain wall will therefore bend sideways, as schematically represented in \fref{fig_d35shear}, a typical shear motion, while the bottom remains clamped. Thus, when an ac electric field is applied through the film, as during PFM measurements, a lateral oscillation of the domain wall occurs, measured as an effective $d_{35}^{DW}$ response.

Within this shear scenario, we can calculate the expected deflection and torsion angles resulting from the purely piezoelectric response, ie. the $d_{33}$ coefficient giving rise to a vertical deflection and the $d_{35}^{DW}$ giving rise to a surface translation below the tip.

\begin{figure}[ht]
\includegraphics[width=\columnwidth]{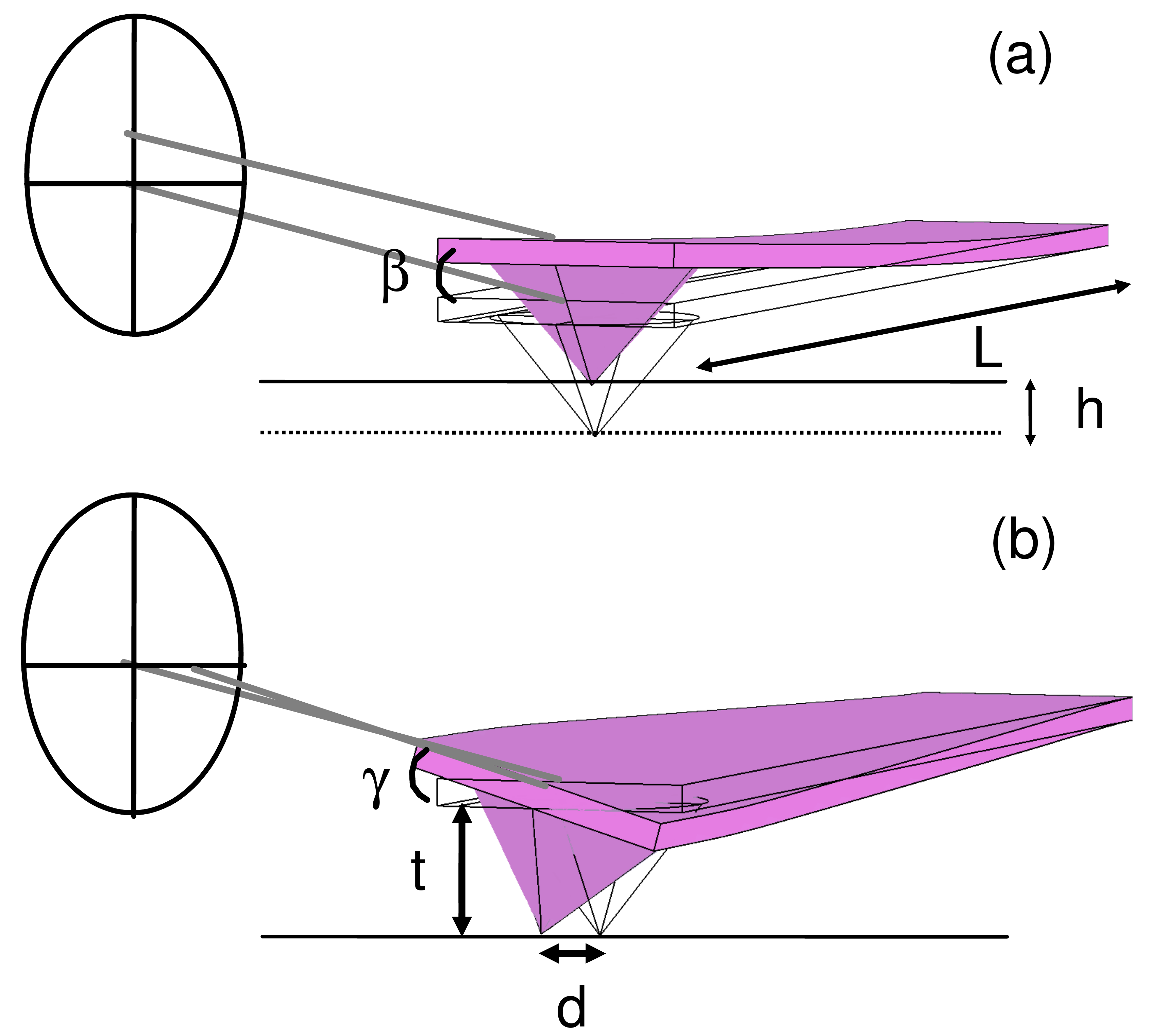}
\caption{Schematic representation of the vertical deflection (a) and lateral torsion (b) of the tip/cantilever system resulting from a change in height of the surface (full and dotted lines in (a)) or in an in-plane displacement of the surface (b) leading to vertical and lateral signals on the photodiodes respectively.}
\label{fig_anglestip}
\end{figure}

\fref{fig_anglestip} gives a schematic picture of the tip/cantilever system when a vertical deflection or a horizontal shear are applied to the tip apex. Taking the same realistic values for the tip/cantilever system as for our \emph{COMSOL} simulations, values of $d_{33}$=75pm/V and $d_{35}^{DW}$=30pm/V at the domain wall \cite{morozovska_resolutionfunction_PFM}, and an applied voltage $V$=1V, we can determine deflection and torsion angles ($\beta$ and $\gamma$) that will lead to vertical and horizontal signals on the photodiodes, respectively, as:
\begin{eqnarray}
\begin{split}
\beta&=\arctan(\frac{h}{L})=\arctan(\frac{2d_{33}\times V}{L})\\
 &=\arctan(\frac{150.10^{-12}}{230.10^{-6}})=3.7.10^{-5}\,^{\circ}
\end{split}
 \label{beta}
\end{eqnarray}
\begin{eqnarray}
\begin{split}
\gamma&=\arctan(\frac{d}{t})=\arctan(\frac{d_{35}^{DW}\times V}{t})\\
&=\arctan(\frac{30.10^{-12}}{20.10^{-6}})=8.6.10^{-5}\,^{\circ}
\end{split}
\label{gamma}
\end{eqnarray}

As can be seen from in this simple calculation, the vertical deflection and torsion angles given by the shear strain scenario are of the same order of magnitude, and would thus result in similar vertical and horizontal difference signals on the photodiode. This conclusion is in agreement with our experimental observations, suggesting that the shear motion of domain walls is a plausible explanation of the observed LPFM response.

More quantitative analysis of the domain wall shear and the measured LPFM signal, taking into account all the non-zero terms of the piezoelectric tensor ($d_{31}$, $d_{33}$, and $d_{15}$)  was carried out using resolution function theory by Morozovska {\it et al}  \cite{morozovska_resolutionfunction_PFM} in several materials, specifically considering the symmetry of the piezoelectric tensor and the breaking of this symmetry at the domain walls. In purely $c$-oriented tetragonal ferroelectric films, for which the $d_{35}$ piezoelectric coefficient is zero by symmetry within a domain in the bulk material. However, as a result of the broken symmetry (the sign change in the $d_{33}$ piezoelectric coefficient on either side of the wall)  local enhancement of the shear response at the exact location of a 180\de domain wall is expected to occur, detectable as an ``effective $d_{35}$ coefficient'' which we have labelled $d_{35}^{DW}$.

To compare our results with the numerical simulations reported by Morozovska \textit{et al.}, 
we determined the full width at half maximum (FWHM) of the LPFM amplitude signal, and the distance 
over which the $d_{33}$ piezoelectric coefficient changes sign (defined as the distance between $\pm$75\% 
of the total VPFM signal $R\cos{\varphi}$, where $R$ and $\varphi$ are the amplitude and phase, respectively).
 For the domain walls shown in \fref{fig_pztpfm}  the LPFM amplitude signal extends over 46$\pm$5 nm, and 
more generally varies between 40 and 70 nm for domain walls at different orientations to the cantilever axis.  
The corresponding distance over which the VPFM signal change sign extends over 62$\pm$3 nm, and more generally
varies between 50 and 90 nm. Following Morozovska, we normalize the values obtained from \fref{fig_pztpfm} by 
the nominal tip radius of 50 nm, giving an LPFM amplitude FWHM of 0.92,  and a distance for the $d_{33}$ sign 
change of 1.24. These ratios are comparable with the values of 0.95 and 0.93, respectively, obtained from the 
numerical simulations. One probable source of discrepancy is the actual tip radius. Studies \cite{bloo_tip_wear,
chung_tip_wear} of the deterioration of an AFM tip in contact mode suggest that even with a few scans the tip 
size increases significantly from its nominal radius. A larger tip radius would lead to lower values of the 
renormalized LPFM amplitude FWHM and distance over which the VPFM signal changes sign, in closer agreement 
with the numerical predictions of Morozovska \textit{et al.} in the latter case.
Furthermore, possible presence of background in the PFM measurement setup could decrease the sharpness of the signal at domain walls, leading to higher measured values.

\section{Other types of materials and domain walls}

From simple elastic considerations, the shear motion due to antiparallel vertical deformation of the material on opposite sides of the domain wall should not be limited to 180\de domain walls in tetragonal ferroelectrics, and we would expect it to occur at every domain wall for which the surrounding domains present at least some component of antagonistic vertical motion. To test this
hypothesis, we therefore measured the VPFM and LPFM response at different types of ferroelectric domain walls in both (111)-oriented and (001)-oriented BFO films, whose more complex ferroelectric structure gave us the opportunity to look at 71, 109 and 180\de domain walls \cite{zavaliche_BFO_domains}.

\subsection{BFO 111}
\begin{figure}[ht]
\includegraphics[width=\columnwidth]{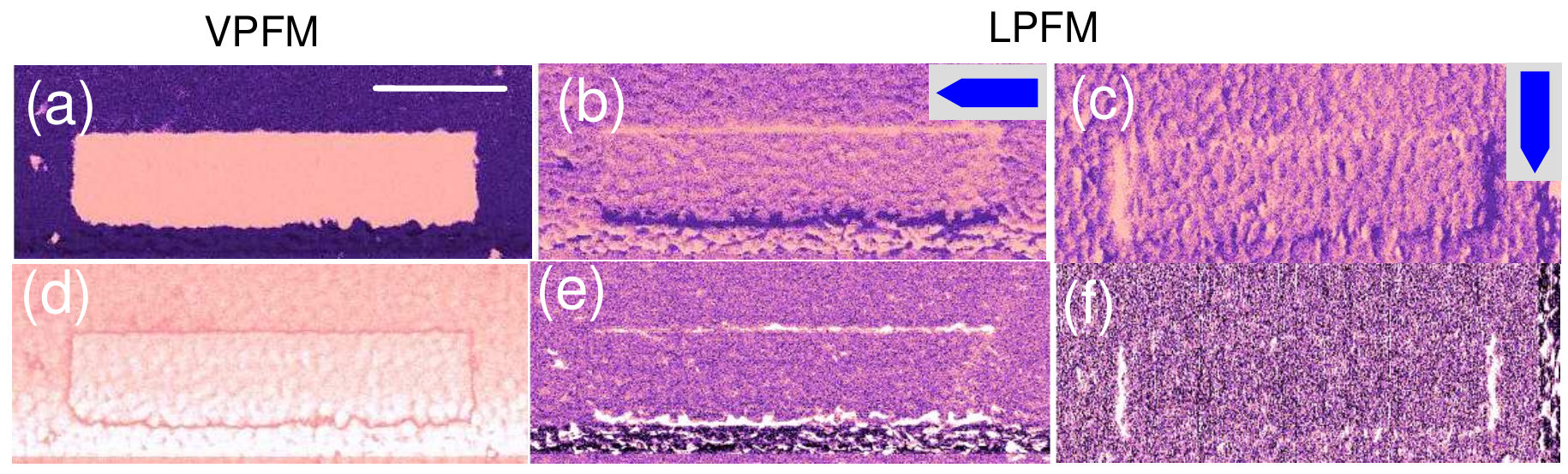}
\caption{(111)-oriented BFO film: VPFM phase (a) and amplitude (d) and LPFM
phase (b)-(c) and amplitude (e)-(f) after writing a rectangle with positive
voltage. In insets of (b)-(c), the orientation of the cantilever during LPFM
measurement is given. The white horizontal bar is 1$\mu$m. The fast scan directions is along the length of the cantilever. (measurements (a) and (d) were made at the same time as (b) and (e).) }
\label{fig_bfo111}
\end{figure}

We carried out PFM measurements on (111)-oriented BiFeO$_3$ thin films deposited on (111) SrTiO$_3$ substrates, presenting a rhombohedral structure \cite{BeaAPL111} with one of the four $<$111$>$ directions being purely out-of-plane and the three others being mainly in-plane.
In the films studied, we find that the vertical component of the polarization is uniformly oriented ``up'', as can be seen in the outer part of the VPFM image in \fref{fig_bfo111}(a), which presents a uniform dark contrast. In this as-grown region, the LPFM phase also presents a relatively homogeneous contrast, although with some small-scale structure, shown in \fref{fig_bfo111}(b) and (c) for two orthogonal cantilever orientations.  These small variations can be readily
correlated with the surface morphology of the film, whose roughness is higher than that of the previously considered PZT, as shown in \fref{fig_roughness}(d), and thus allows surface contributions to appear in LPFM \cite{peter_PFM_shapeeffects}. However, outside of these small features, no other LPFM signal is apparent in the as-grown film, and the mid-scale contrast
suggest that in fact there is very little lateral response, and thus no contribution from an in-plane component of the polarization.  The as-grown films are therefore monodomain, with a purely out-of-plane polarization directed along the corresponding $<$111$>$ axis.

In a selected region of the film, we wrote rectangular domains with a positive tip voltage, one of which is shown in \fref{fig_bfo111}. As expected, the VPFM phase contrast in the written region is now bright corresponding to a ``down'' polarization (see \fref{fig_bfo111}(a)), and at the domain wall, a decrease in VPFM amplitude is observed (\fref{fig_bfo111}(d)). In the LPFM images of
\fref{fig_bfo111}(b) and (c), no change is observed inside the written domain (although once again there are small-scale features related to the morphology of the sample surface). We can thus conclude that the ``up'' polarization in the as-grown region and ``down'' polarization in the written region are separated by 180\de domain walls. In the LPFM phase, as for PZT, we observe dark (on the left of the cantilever) and bright (on its right) contrast at the position of these
domain walls when they are perpendicular to the cantilever: in \fref{fig_bfo111}(b), this alternate contrast is on the longer domain walls while in \fref{fig_bfo111}(c), it is on the shorter ones. Again, similarly to PZT, this contrast is accompanied by a strong increase in the LPFM amplitude, as can be seen in \fref{fig_bfo111}(e) and (f).

Thus, in this non-tetragonal material, we also observed a similar lateral PFM signal at the position of the 180\de domain walls.
\subsection{BFO 001}
\begin{figure}[ht]
\includegraphics[width=\columnwidth]{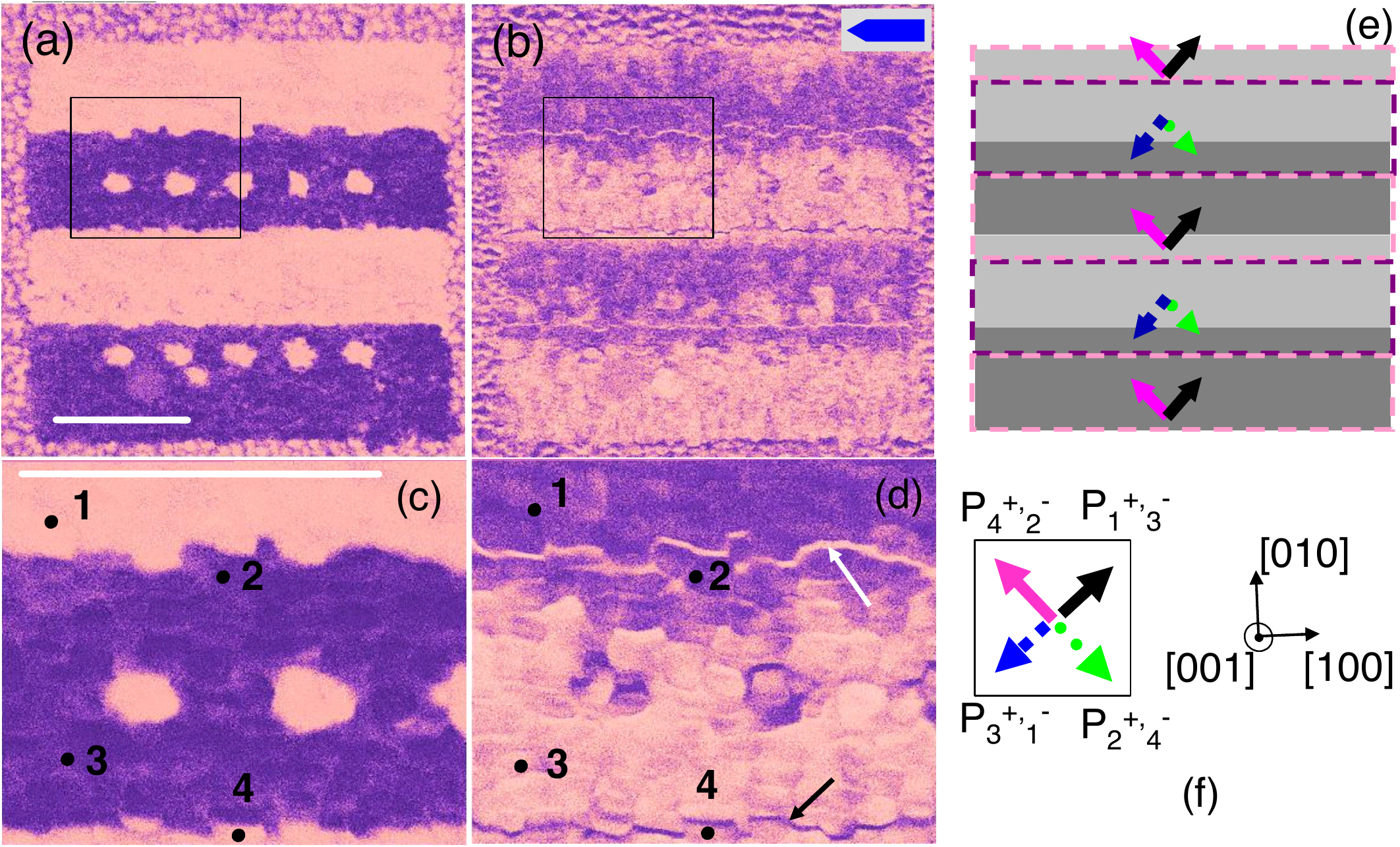}
\caption{(001)-oriented BFO film.  (a), (c), VPFM (b), (d) and LPFM phases after writing large rectangles with negative (dark color in VPFM) and positive (bright color in VPFM) voltage with slow scan axis along [0$\overline{1}$0]. Circular nanodomains were then written with voltage pulses. (c)-(d) are a zoom in the region indicated by the rectangle in (a)-(b). For points, numbers and arrows, see text. (e) gives a schematic representation of the polarization orientation
of (a)-(b): light and dark grey correspond to down and up out-of-plane component of the polarization respectively and the arrows indicate its in-plane orientation (two adjacent arrows indicate the mix of components). (f) gives crystalline axis orientation and in-plane polarization variants orientation, + and - gives the out-of-plane component. The white bar is 1$\mu$m.}
\label{fig_bfo001}
\end{figure}

We then studied the more complex (001)-oriented BFO films. In this case, the eight possible variants of the polarization present both in-plane and out-of-plane components. As can be seen in \fref{fig_bfo001}(a) and (b), in the as-grown region (outer part), both VPFM and LPFM phase images show a multidomain pattern. In this film, all components of the polarization are thus present.

We then wrote rectangular domains with positive and negative voltage, scanning the tip with the slow scan axis along [0$\overline{1}$0], and subsequently patterned an array of 5x5 circular nanodomains by applying positive voltage pulses to the stationary tip at programmed positions.  As shown in VPFM phase measurements in \fref{fig_bfo001}(a) the regions corresponding to both
``up'' (dark contrast) and ``down'' (bright contrast) out-of-plane polarization components were successfully written. As can be expected, the circular nanodomains, written with positive voltage, are visible only when present in an ``up''-polarized region. In the corresponding LPFM phase image of \fref{fig_bfo001}(b), which allows us to determine the component of the polarization along the [010] direction, we note that the in-plane component of the polarization has also been modified by the application of an electric field to the scanning tip. This in-plane polarization switching has already been observed in our previous study \cite{Beaswitching} and in another recent study \cite{Balke_PFM} and has been attributed to the very small horizontal electric field provided by the tip combined with the tip motion. If we now look more closely in the written regions (\fref{fig_bfo001}(c,d)), we see that within a domain that presents only one out-of-plane component of the
polarization (either dark or bright contrast in \fref{fig_bfo001}(c)), the in-plane polarization component is not homogeneous, as can be seen in \fref{fig_bfo001}(d). For example, within the region showing a dark contrast in the VPFM image, the LPFM contrast is either bright  or dark (see region 2 vs. 3 in \fref{fig_bfo001}(c) and (d)). Similarly, in the region showing bright contrast
in VPFM, different LPFM contrasts can be observed (see regions 1 and 4 in \fref{fig_bfo001}(c) and (d)). The LPFM image taken after rotation of the sample by 90\de (not shown) indicates that along the [100] direction, the written domain present both polarization components. We can thus fully determine the orientation of the polarization in the different domains (see \fref{fig_bfo001}e), the arrow orientation giving the in-plane component and the gray level giving the out-of-plane component of the polarization. Thus by crossing the domain wall between regions 1 and 2, the polarization goes from $P_1^{-}$ to $P_3^{+}$ or $P_2^{+}$ or from $P_4^{-}$ to $P_3^{+}$ or $P_3^{+}$, as defined in \fref{fig_bfo001}f. Between regions 1 and 2, one thus encounters either 71 or 109\de domain walls with opposite out-of-plane components of the polarization. A similar analysis may be done between regions 3 and 4.

 If we consider specifically the regions around the domain walls separating different out-of-plane polarization components, we note that the LPFM contrast is  the same on either side of the domain walls and that the in-plane polarization component along the [010] direction is thus not changed. However, at the exact position of this out-of-plane domain wall a sharp LPFM contrast reversal is observed, ie. a thin dark line is visible in the bright contrast region (see dark arrow in
\fref{fig_bfo001}(d)) and a vice-versa, a thin bright line may be seen in the dark contrast region (see white arrow in \fref{fig_bfo001}(d)).

\begin{figure}[ht]
 \includegraphics[width=\columnwidth]{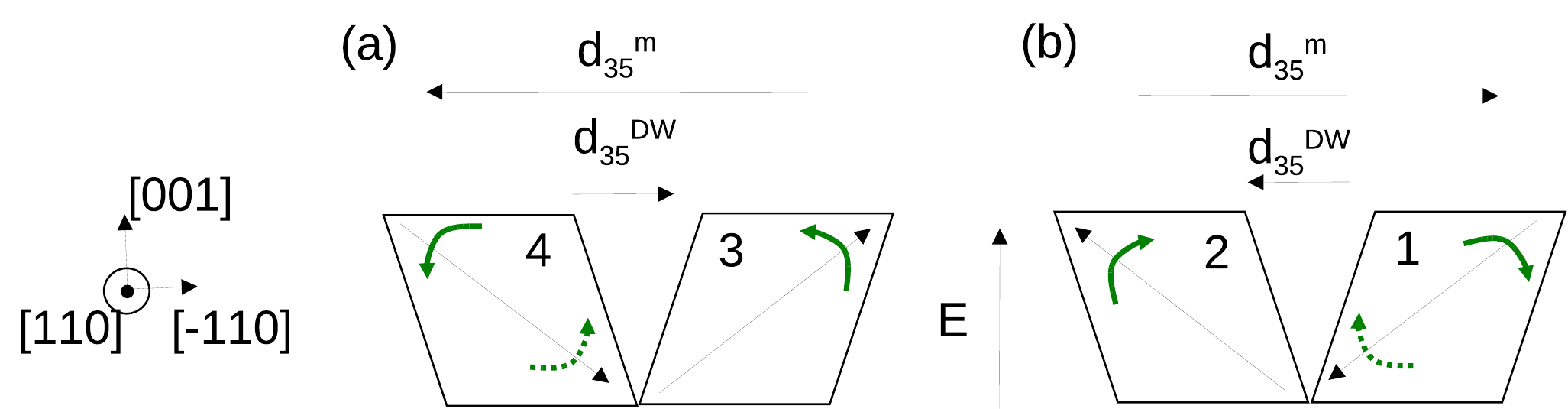}\\
\caption{Schematic representation of the unit cells and of the polarization projected in the (110) plane in regions 1 to 4 of \fref{fig_bfo001}c-d (see numbers):
at the domain wall between 3 and 4 (a) or 1 and 2 (b). The effect of a field oriented along [001] is given by the green curved arrows and the resulting shear displacements at the surface due to the bulk, $d_{35}^m$, or to the domain wall, $d_{35}^{DW}$, are represented by straight arrows along either [010] or [0$\overline{1}$0].}
\label{mechd35mDW}
\end{figure}
This sharp opposite contrast line at the domain wall in LPFM response may be explained by a similar shear contribution to that observed in PZT and (111)-BFO, as can be seen in \fref{mechd35mDW}. Under the action of an electric field applied along the out-of-plane [001] direction, the polarization tends to rotate to align with the field, as shown by the green curved arrows in
\fref{mechd35mDW}. In regions 3 and 4 (\fref{mechd35mDW}(a)), due to the non-zero bulk $d_{35}$ coefficient, henceforth referred to as $d_{35}^m$, the surface  shears towards [1$\overline{1}$0]. However, at the domain wall separating these regions, the local $d_{35}^{DW}$ coefficient leads to a shear along [$\overline{1}$10] since region 3 will contract and region 4 expand
in the in-plane direction (ie. in both [110] and [1$\overline{1}$0] directions) as a result of their antiparallel vertical deformation. This opposite lateral motion of the surface in the domain and at
the domain wall will therefore give opposite LPFM contrast. The highly enhanced amplitude of the shear strain response ($d_{35}^{DW}$) at the precise position of the domain wall \cite{morozovska_resolutionfunction_PFM} allows this signal to locally dominate over the bulk response due to the uniform in-plane polarization component ($d_{35}^m$).

It is important to note here that the lateral extension of the domain wall response due to $d_{35}^{DW}$ is smaller in (001)-BFO as compared to (111)-BFO or PZT. The important difference is that in the former, as we have just shown, a competition between the bulk domain and the domain wall responses occurs, which is not the case for (111)-BFO or PZT, since these two materials do not present a bulk $d_{35}$ response. In all cases, the value of $d_{35}^{DW}$ decreases with the distance between the tip and the domain wall \cite{morozovska_resolutionfunction_PFM}. In
PZT and (111)-BFO, where no bulk contribution is present, the shear response due to $d_{35}^{DW}$response will be measurable until it decreases below instrumental resolution. In (001)-BFO the domain wall response will be more rapidly cancelled out by the bulk domain contribution $d_{35}^m$ leading to a much sharper feature.

A similar analysis may explain the opposite contrast between regions 1 and 2 (bright contrast in \fref{fig_bfo001}d) and the separating domain wall (dark contrast shown by the white arrow in \fref{fig_bfo001}d) as shown in \fref{mechd35mDW}(b).

In (001)-BFO thin films, a shear signal similar to the one at 180\de domain walls has thus been observed in 71 and 109\de domain walls presenting a reversal of the out-of-plane component of the polarization.

\section{Conclusions}
In this paper we have explored in PZT and BFO thin films the lateral PFM signal observed at 180\de domain walls separating regions with antiparallel out-of-plane polarization. due to a shear response. We have determined that neither electrostatic effects nor a tilting of the film surface can satisfactorily explain the observed phenomenon.  Rather, the shear response is specifically allowed  as a result of the breaking of bulk symmetry by the sign change of the piezoelectric deformation across the domain wall under the action of an applied vertical electric field.This translates into a shear movement of the domain wall, which could be of significant interest for applications requiring the horizontal propagation of a surface deformation, such as nanomechanical transducers in surface acoustic wave devices \cite{kumar_SAW}. We then demonstrated that this scenario can be successfully generalized to other types of domain walls, which show a similar shear response as long as the regions on either side present anti-parallel out-of-plane polarization components.  In materials such as BFO, the shear response competes with the signal due to a uniform in-plane polarization component across such domain walls, and has to be carefully considered during PFM analysis of these materials. Let's stress that, as this phenomenon is due to elastic response of the material, it may change in amplitude and/or sign  for materials with different elastic properties \cite{morozovska_resolutionfunction_PFM}.

\begin{acknowledgements}
The authors thank E. Soergel, K. Bouzehouane, S. Fusil and J.-M. Triscone for helpful discussions, F. Guy and S. Gariglio for the growth and characterization of the PZT thin films, and M. Lopes for technical assistance. This work was supported by the Swiss National Science Foundation through the NCCR MaNEP and Division II, and by the European Commission STREP project MaCoMuFi. H. B\'ea was supported by Bourse d'excellence from the University of Geneva.
\end{acknowledgements}
\bibliographystyle{prsty}

\end{document}